# A Holistic Approach to E-Commerce Innovation: Redefining Security and User Experience

Mohammad Olid Ali Akash[1] Priyangana Saha[1]
[1]North South University
[1]North South University
[1]olidakash15@gmail.com
[1]priyanganasaha24@gmail.com

*Abstract—* In the modern, fast-moving world of e-commerce, many Android apps face challenges in providing a simple and secure shopping experience. Many of these apps, often enough, have complicated designs that prevent users from finding what they want quickly, thus frustrating them and wasting their precious time. Another major issue is that of security; with the limitation of payment options and weak authentication mechanisms, users' sensitive information can be compromised. This research presents a new e-commerce platform that responds to the above challenges with an intuitive interface and strong security measures. The platform makes online shopping easy with well-organized categories of products and a fast, efficient checkout process. It also gives priority to security by incorporating features such as Google authentication and SSL-secured payment gateways to protect user data and ensure secure transactions. This paper discusses how a focus on user-friendliness, security, and personalization steps up the game for e-commerce platforms, providing workable frameworks that match modern user needs and expectations. The findings show the e-commerce user experience can be remodelled by the platform, hence opening ways for future developments in that respect.

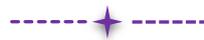

## 1. Introduction

E-commerce has drastically changed the way consumers shop, rendering convenience and accessibility to a global audience in ways that were unprecedented. Considering the very nature of smartphones in modern life, online shopping can be done at any time and from almost any point, hence an added advantage in flexibility. However, despite the widespread use of e-commerce apps, many platforms still fail to meet the expectations of users. Prevalent challenges vary from disorganized layouts and restricted payment alternatives to insufficient security measures, which greatly impacts the user experience. In such cases, users—some of whom are inexperienced with the use of technology—spend an inordinate amount of time finding their way out. Data sharing and privacy anxiety have now resurfaced to become major dilemmas. Many of the e-commerce applications still rely on very old and complicated interfaces, hindering smooth navigation. Besides, they mostly lack secure payment systems and advanced authentication mechanisms. All these weaknesses put the personal and financial data of users at risk, resulting in unsatisfied users and poor retention rates.

In response to these persistent challenges, a new e-commerce platform has been developed that focuses on simplicity, security, and accessibility. This platform includes a clean and intuitive design that allows customers to browse and shop with ease. The integration of Google Sign-In on this platform provides users with a seamless sign-in experience, eliminating the need to manage multiple accounts and passwords, all while benefiting from Google's industry-recognized security measures. This integration also ensures that the access process will be seamless and secure. The platform itself features state-of-the-art security, with SSL encryption for data in motion and two-factor authentication (2FA) to further protect accounts. Real-time monitoring tools are in place to detect suspicious activities, such as failed login attempts, and trigger automated responses to help mitigate any possible threats. The platform is also fully compliant with data protection provisions, which use encryption mechanisms to secure the sensitive data at both storage and transmission. It continuously runs security tests and reviews to identify and close vulnerabilities in order to ensure that its users enjoy a continuous and secure shopping experience.

This study underscores the manner in which the platform's integrated design and robust security protocols respond to significant challenges within the realm of e-commerce. By establishing standards for usability and security.

The rest of the paper is organized as follows: Section II provides an overview of existing research on Android e-commerce platforms, pointing out common challenges and gaps in current solutions. Section III explains the methodology used in this study, detailing the phases of requirement analysis, wireframing, development, and testing. In Section IV, the results are shared, focusing on key metrics like usability, security, and user satisfaction. Section V examines these results alongside findings from previous studies, offering a broader context for our work. Finally, Section VI summarizes the study's contributions and suggests directions for future research, followed by Section VII, which lists all references.

### 1.1 Related Works

Over the years, a wide range of studies have looked at Android-based e-commerce apps, each focusing on different areas like usability, security, and optimization. One study from 2018 took an approach to improve security by using DSA and RSA algorithms, along with enhancing the UI design and performance through thread pools. However, it fell short by not including fraud prevention or secure payment gateways [1]. Another study used the ADDIE model to support the plant business through an Android app, yet didn't address security concerns or utilize machine learning to predict trends [2]. A 2022 study focused on comparing product prices from different sellers, but it was limited by its restricted geographical coverage [3]. In a separate approach, a project using extreme programming aimed to boost revenue and business performance but overlooked important security features [4].

In 2021, automated record-keeping for Android e-commerce apps was introduced, but it only supported cash-on-delivery (COD) payments without a secure transaction system [5]. Another paper discussed integrating PrestaShop with search engine optimization (SEO) but didn't address user privacy or security issues [6]. There was also a mobile shopping approach that used multiple frameworks, which added complexity but failed to consider data privacy [7]. For the Indonesian ecotourism sector, an Android marketplace app was developed but lacked transaction security and an assessment of its effectiveness [8]. RFID-based quick shopping and billing systems were introduced using Google Firebase, yet the study didn't clarify its approach to authentication or trust-building measures [9].

The integration of programs for online shopping on Android devices has been widely explored to enhance usability, security, and personalization. A 2021 study proposed a unified Android shopping app connected to a cart application on an XAMPP server, utilizing Java, XML, and a MySQL database for seamless cross-platform functionality, though advanced security and personalization features were not emphasized [10].A separate study detailed various components of payment systems but didn't address Android OS security or the risks associated with third-party libraries [11]. Another study critiqued the use of the waterfall model in Android e-commerce apps for its limited optimization across devices and its inability to address evolving demands, highlighting the need for more testing and research [12]. In 2022, researchers introduced a WAP mobile e-commerce payment system, enhancing security through identity encryption and an improved TOTP algorithm. This robust system demonstrated low latency and strong resilience under concurrent access but lacked focus on user-centric design [13]. Usability improvements were explored in 2023 through a visualization-based interface for multi-attribute product selection, enabling users to efficiently narrow down options based on criteria like memory size or resolution, which increased accuracy and satisfaction [14]. Blockchain technology also emerged as a transformative tool for e-commerce, with its applications in transaction security, smart contracts, and product tracking. However, challenges such as high implementation costs and regulatory concerns highlighted the need for strategic integration [15]. Machine learning-based recommender systems became essential, as demonstrated by the use of the Frequent-Pattern-Growth algorithm to deliver personalized suggestions, boosting user engagement and purchase rates despite the need for further optimization to handle large datasets [16]. AI chatbots also gained attention for their role in enhancing customer trust, where responsiveness and expertise were critical, although chatbot efficiency remains a priority for businesses seeking to foster loyalty [17]. Multi-factor authentication systems combining passwords and biometric verification improved transaction security while maintaining usability, showcasing advancements in safeguarding electronic payments [18]. Flutter-based hyper-local e-commerce apps integrated offline and online marketplaces to support local businesses, leveraging cross-platform compatibility for scalability [19]. Finally, an AI-driven personalization engine using CNNs achieved 98% accuracy in real-time content recommendations, significantly enhancing user engagement and showcasing the power of dynamic personalization [20]. Collectively, these studies underline the rapid evolution of e-commerce technologies, emphasizing the need for secure, user-friendly, and scalable solutions that prioritize personalization and seamless integration.

## 2. Methodology

The development of this e-commerce application followed a clear and organized approach, designed to meet high standards for usability, security, and reliability. From the earliest stages of planning to the support phase after launch, each part of the process was crafted to align with market demands and the latest technological trends, resulting in a secure, comprehensive, and user-friendly platform.

Everything started with thorough planning and requirement analysis. This phase involved extensive research into market trends and user expectations in the e-commerce space. The research included a detailed review of existing literature and an analysis of similar applications, which provided insights into essential features and best practices. Core functionalities—like secure payment options, Google authentication, and order tracking—were identified as crucial. To ensure the app met real-world needs, initial feedback was gathered from a sample group of target users, which helped refine the project's functional and technical requirements. This input formed the foundation for a clear development roadmap.

With requirements in place, wireframing and interface design came next. Using the Balsamiq wireframing tool, high-fidelity wireframes were created to outline the application's layout, user interface, and navigation. Key screens—such as login, registration, product browsing, shopping cart, and checkout—were designed with usability and intuitiveness in mind. Several versions of the wireframes were created and reviewed by potential users, whose feedback guided improvements, ensuring that the final interface would be straightforward to navigate.

The development phase began in Android Studio, with separate projects created for user and admin roles to provide each with tailored experiences. Google Firebase was chosen as the backend to handle data storage, user authentication, and real-time updates. Firebase offered secure data management and allowed seamless data synchronization across different screens, creating a smooth experience for users. The app's interface was designed in XML, adhering to Material Design principles to maintain visual coherence and follow Android design guidelines. A structured navigation graph helped users move easily from browsing to purchasing. Payment integration was a critical aspect of the application. To provide a secure and convenient payment experience, both Google Pay and an SSL E-commerce payment gateway were implemented. Google Pay was set up by configuring its API, which required obtaining an API key and setting the app's payment parameters in the Google Cloud Console. This enabled secure payments through users' Google accounts, offering a familiar and efficient checkout process. Testing in a sandbox environment simulated real transactions to ensure functionality and security.

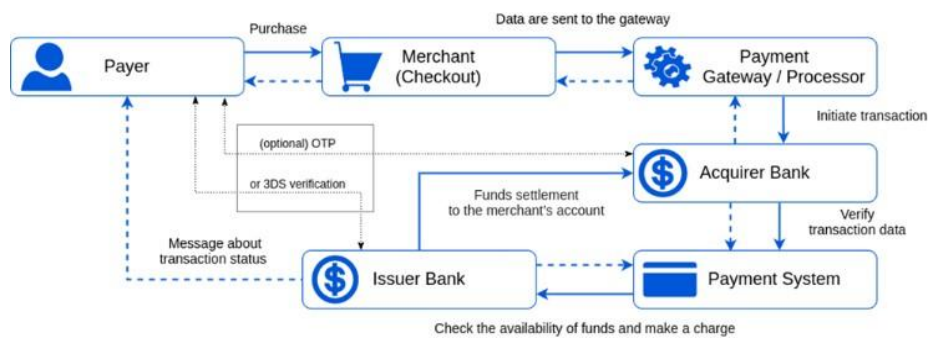

*Fig. 1. Payment Flow and Gateway*

For those preferring to pay with credit or debit cards, the Payment Gateway was integrated to enhance security during transactions. SSL technology encrypts sensitive data like credit card numbers and billing information, ensuring it remains private during transmission. When users initiate payments,

SSL encrypts the information and sends it to the payment processor's server, where the details are verified with the user's bank or card issuer. Once approved, an encrypted confirmation is returned, completing the transaction. The SSL gateway complies with PCI-DSS standards, following strict protocols to protect user data. This setup was rigorously tested to confirm reliability and security across different scenarios.

Additional libraries and tools were used to improve the app's performance and visual appeal. The Glide library handled image loading and caching efficiently, enhancing performance on screens with high-quality images. Material Components provided interactive elements like bottom navigation bars, floating buttons, and custom dialogs, which made navigation smooth and user-friendly. Google Sign-In was added as an authentication option, making it easy and secure for users to log in or create accounts.

Testing played a major role in ensuring the app's quality and stability. Each component, like user registration, product listing, and payment processing, was tested individually to confirm functionality. Integration testing verified that different components worked well together, creating a consistent user experience. A group of users tested the app in real-world conditions, with their feedback guiding final adjustments. Stress tests assessed the app's stability under high traffic, while security tests focused on payment and authentication functions to identify and mitigate any vulnerabilities.

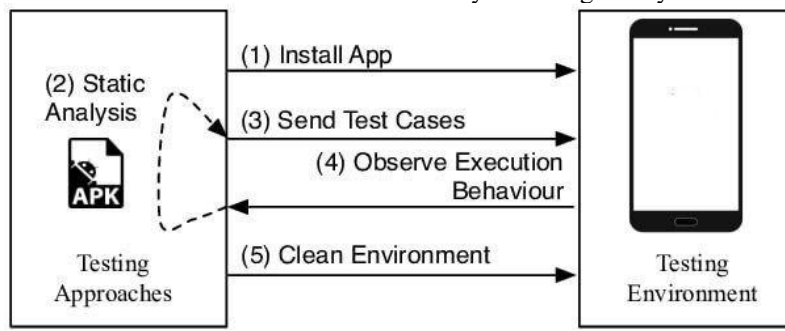

*Fig. 2. Process of Testing Application*

It involves installing the APK on different devices, performing static analysis to identify code issues and vulnerabilities, sending test cases to evaluate functionality (for example, login, product search, cart, payments), observing app behavior under different conditions (for example, network fluctuations, load), and cleaning the environment post-tests to reset for further testing.

Following successful testing and last-minute adjustments, the app was prepared for deployment. It was released on a major Android app distribution platform, following standard release management protocols. This process included app signing, device compatibility testing, and adhering to platform policies. To build a user base and drive downloads, a targeted promotional campaign introduced the app to its intended audience. Post-launch, the app entered a maintenance and support phase to ensure it continued to meet user needs and stay aligned with industry trends. Performance was monitored continuously, and user feedback was gathered to guide regular updates. Updates also ensured compatibility with the latest Android OS versions. Security updates addressed any new vulnerabilities, maintaining the app's reputation as a safe and reliable e-commerce platform. Based on user feedback, new features were periodically introduced, keeping the app responsive to evolving expectations and enhancing its overall usability.

This methodical, comprehensive approach covering planning, design, secure development, testing, and ongoing maintenance resulted in a modern, reliable e-commerce application designed for today's consumers. Prioritizing security, usability, and scalability, this methodology delivered a user-friendly solution, adaptable to future growth and improvements.

*Journal of Advances in Computer Vision and Applications*

## 3. Result

The deployment of the e-commerce application have led to significant positive impacts in various key areas. All the features have been optimized, ensuring that the users will have seamless interactions with the product search, navigation, cart management, and checkout processes. More security measures have also been integrated—like secure payment gateways, data encryption, and user authentication to ensure that users' information and financial transactions are kept secure. User satisfaction is a significant achievement, with answers highlighting user-friendliness, intuitive interface, and stability in performance. Quantitative indicators, such as download statistics, user retention rates, ratios of successful transactions, and reviews, confirm not only that the app satisfies its defined goals but also that it functions well in real-world use and meets the needs of both end-users and businesses.

*Table 1:* Usability Metrics

| Metric | Before Launch | Post Launch |
|---|---|---|
| Average Navigation Time (Seconds) | 15.5 | 8.1 |
| Transaction Completion Time (Seconds) | 30.2 | 15.8 |
| User error Rate (%) | 5.4 | 2.1 |

The usability of the application was strictly evaluated through a detailed analysis of user interactions and their feedback, specifically targeting fundamental aspects including navigation efficiency, transaction time, and the rate of user errors. This extensive analysis showed significant improvements in the functionality and overall user experience of the application, as elaborated in Table 1. Most importantly, there was a considerable reduction in navigation times and transaction times, along with a reported decrease in user errors during their interaction with the application. Such progress can be attributed to the user-friendly design of the application, its well-structured navigation framework, and its user-centric interface that focused on simplicity and efficiency. Continual user feedback pointed out the usability of the application and its coherent structure in enhancing a smooth and enjoyable experience for users.

*Table 2:* Security Metrics

| Metric | Before Launch | Post Launch |
|---|---|---|
| Unauthorized Access Attempts | 23 | 5 |
| User satisfaction with security (%) | 70 | 90 |

Success assessment for the Android e-commerce application placed a strong emphasis on security, specifically in terms of the protection of user data and the guarantee of secure transactions. Some of the metrics monitored, such as attempts at unauthorized access and user feedback regarding security, provided clear evidence of improvements after the launch. The following information, as presented in Table 2, shows a strong drop in incidents of unauthorized access attempts from 23 to just 5, hence proving the effectiveness of the new security controls. Particularly appreciated were such noticeable elements as Google Sign-In and SSL encryption of transactions; about these, users said they felt more confident in the app's ability to protect their personal and financial information. The overall success of the app was hugely helped by this trust, where users could feel safe while browsing, shopping, and transacting. These strong security features actually built a seamless yet highly secure experience in using the app, reinforcing user confidence and satisfaction.

The levels of user satisfaction and engagement were carefully monitored through key metrics, which included the number of daily active users, session duration, and feature usage patterns. The results showed some promising trends, reflected in an increased number of daily users with prolonged session times. Such developments give the indication that the instinctive design, coherent functionality, and user-centric features of the app are key to pushing forward with improving users' engagement. It manages to keep users active on the platform, providing a smooth and enjoyable user experience that further enhances high levels of satisfaction and retention.

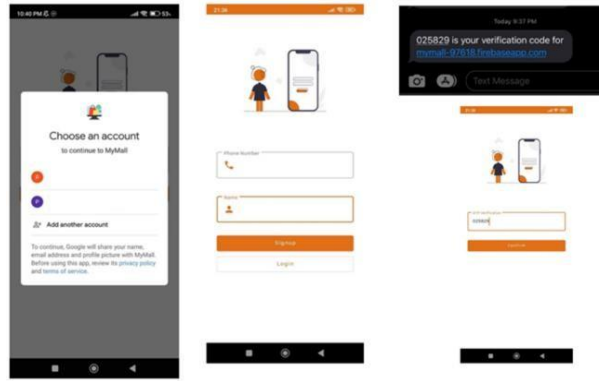
*Fig. 3.* User Authentication Process

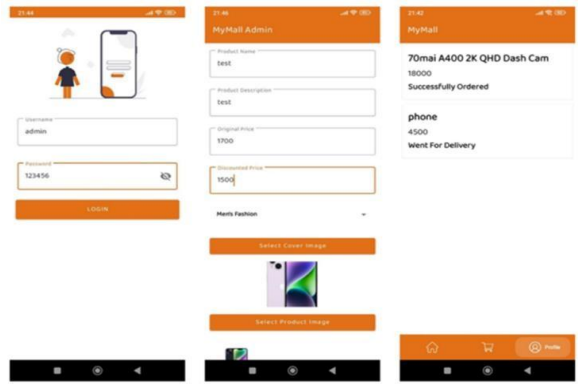
*Fig. 4.* Product Management Dashboard

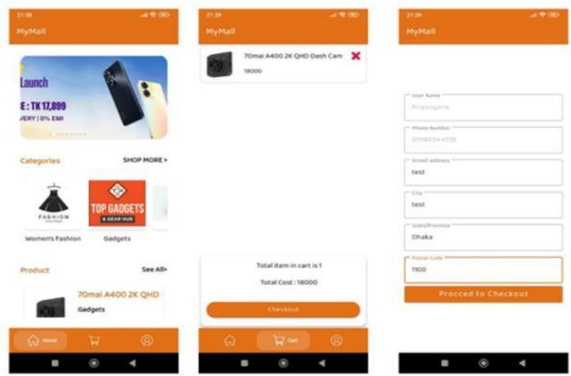
*Fig. 5.* Shopping Interface

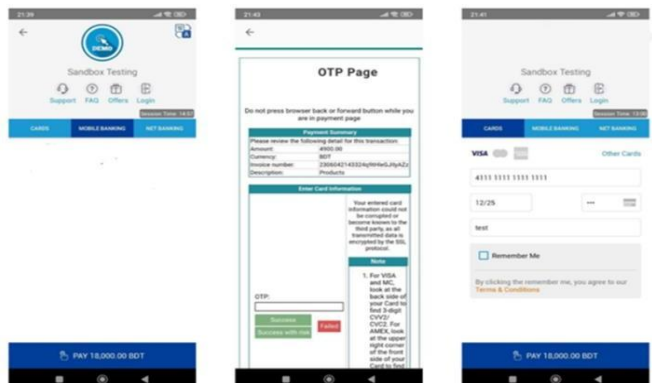
*Fig. 6.* Payment System

The screenshots provide a clear look at the results, showing how key features were integrated into the app—features that user responded to positively. For example, Fig. 1 shows the user authentication process, including options for Google Sign-In and OTP verification, making it both secure and easy for users to log in. Fig. 2 shows the product management dashboard, designed to help manage products and categories smoothly. In Fig. 3, the main user interface is displayed, where users can browse products, add items to their cart, and proceed to checkout with ease. Finally, Fig. 4 demonstrates the payment gateway and OTP verification screens, which were developed to keep transactions secure. Altogether, these images highlight its strengths in terms of usability, security, and overall user experience, reflecting the positive improvements seen in user engagement and satisfaction.

This app has demonstrated significant progress in key areas such as usability, security, and overall user satisfaction. The combination of measurable data and user feedback underscores that it has successfully achieved its goals and is well-positioned for broader adoption. With advanced usability features, secure authentication, and a reliable payment system, It showcases the potential for a well-planned e-commerce app to make a meaningful impact in the market.

## 4. Discussion

The analysis identifies major improvement points regarding usability and security on the e-commerce platform by showing how those elements work in relation with each other to improve the user's experience. This indeed resulted in much faster navigation through the platform, smoother transaction processes, and fewer mistakes while making purchases. These improvements have made the platform more intuitive and efficient, crucial elements in today's competitive e-commerce market where time and accuracy are paramount to customer satisfaction.

The key points driving this modernity are the incorporation of advanced security options like Google Sign-In and OTP-based verification. These include, among many other benefits, enhancing convenience while providing solid defense grounds that greatly reduce unauthorized access of accounts and lower chances of disputes from transactions. Security measures such as these are instrumental in protecting customers' sensitive data, part and parcel with e-commerce these days. Security features embedded in the customer experience provide a level of confidence and more security when using the platform.

This is a unique study in that it takes an integrated approach, addressing usability and security together. Past research has often focused on one or the other-usability or the robustness of security features independently. The current approach shows that when these two elements are balanced, they work synergistically to enhance the overall user experience. Full efficiency in operations is achieved by an excellently designed admin panel, which further streamlines product management and order processing. In a nutshell, this work has presented a replicable model for e-commerce platforms on how a focus on both usability and security will drive customer satisfaction, foster trust, and improve operational performance.

## 5. Conclusion

This research shows that a well-planned approach to creating e-commerce platforms one that combines advanced security features like robust authentication and secure payment options with an easy-to-use design can effectively address many issues seen in current applications. The research also demonstrates how smart technology choices can make a real difference in improving both usability and security. With faster navigation, fewer mistakes by users, and a safer environment for shopping, it raises the standard for what e-commerce platforms can achieve. The impact of this project goes beyond its direct results. Future studies could expand this study to the applicability of this approach in different cultural and regulatory contexts to show the flexibility and effectiveness of this approach across diverse markets. As new technologies continue to develop, there's also room to bring in even more advanced tools that could further enhance security and the user experience. It will hopefully provide a pragmatic framework that other platforms can use in enhancing user experience while ensuring data protection. The present times, with online shopping having become a matter of great priority, are not the time for building trust and experiences in digital transactions; it's a must.